\documentclass[twocolumn,prl, twocolumn,superscriptaddress, nofootinbib, longbibliography]{revtex4-2}

\usepackage[T1]{fontenc}
\usepackage[utf8]{inputenc}
\usepackage{color}
\usepackage{amsmath}
\usepackage{amssymb}
\usepackage{graphicx}
\usepackage{wasysym}
\usepackage[pdfusetitle,
 bookmarks=true,bookmarksnumbered=false,bookmarksopen=false,
 breaklinks=false,pdfborder={0 0 0},pdfborderstyle={},backref=false,colorlinks=true]
 {hyperref}
\hypersetup{
 citecolor=blue,urlcolor=blue}

\makeatletter

\newcommand{\noun}[1]{\textsc{#1}}

\makeatother

\begin{document}
\title{Relaxing towards generalized one-body Boltzmann states }
\author{Sheng-Wen Li}
\email{lishengwen@bit.edu.cn}

\affiliation{Center for Quantum Technology Research, and Key Laboratory of Advanced
Optoelectronic Quantum Architecture and Measurements, School of Physics,
Beijing Institute of Technology, Beijing 100081, People’s Republic
of China}
\author{Ning Wu}
\email{wunwyz@gmail.com}

\affiliation{Center for Quantum Technology Research, and Key Laboratory of Advanced
Optoelectronic Quantum Architecture and Measurements, School of Physics,
Beijing Institute of Technology, Beijing 100081, People’s Republic
of China}
\begin{abstract}
Isolated quantum systems follow the reversible unitary evolution;
if we focus on the dynamics of local states and observables, they
exhibit the irreversible relaxation behaviors. Here we study the local
relaxation process in an isolated chain consisting of \emph{N} three
level systems. Though the entropy of the full many body state keeps
a constant, it turns out the total correlation of this system approximately
exhibits a monotonically increasing behavior. More importantly, a
variation analysis shows that, the total correlation entropy would
achieve its theoretical maximum when each site stays in a generalized
one-body Boltzmann state, which is not solely determined by the energy
but also depends on the spin value of each onsite level. It turns
out such a theoretical correlation maximum is highly coincident with
the result obtained from the exact time dependent evolution. In this
sense, the total correlation entropy well serves as an indicator for
the dynamical irreversibility of the nonequilibrium relaxation in
this isolated system.
\end{abstract}
\maketitle
\noindent \emph{Introduction} - Open systems would be thermalized
to be a thermal state having the same temperature with its reservoir.
However, the thermalization process for isolated many body systems
is tricky to define \citep{mackey_dynamic_1989,landi_irreversible_2021},
since the reversible unitary/Liouville dynamics guarantees the entropy
of isolated quantum/classical systems never changes \citep{hobson_irreversibility_1966,huang_statistical_1987,swendsen_explaining_2008}. 

For an isolate gas with weak particle collisions, focusing on the
one-particle probability distribution function (PDF), Boltzmann derived
a transport equation and proved that the entropy of this one-particle
PDF always increases until it finally reaches the Boltzmann-Maxwell
distribution (the Boltzmann \emph{H}-theorem) \citep{boltzmann_weitere_1872,huang_statistical_1987,chliamovitch_kinetic_2017}. 

Regardless of the dynamical process, focusing on the observable expectations
in the equilibrium state long after relaxation, the ensemble theory
was established based on the statistics interpretation, which forms
the foundation of statistical physics \citep{gibbs_elementary_1902,boltzmann_uber_1877,ueda_quantum_2020}.

To understand the connection between the reversible microscopic dynamics
and the statistical ensemble descriptions, the eigenstate thermalization
hypothesis (ETH) provides an inspiring sight \citep{deutsch_quantum_1991,srednicki_chaos_1994,rigol_thermalization_2008,polkovnikov_colloquium:_2011,ueda_quantum_2020,dalessio_quantum_2016}.
Based on the ergodicity hypothesis, the state of long time average
is considered (\emph{diagonal ensemble} \citep{rigol_breakdown_2009}),
\begin{equation}
\bar{\boldsymbol{\rho}}_{\text{\textsc{d}}}\equiv\lim_{T\rightarrow\infty}\frac{1}{T}\int_{0}^{T}dt\,\hat{\boldsymbol{\rho}}(t)=\sum P_{k}\,|E_{k}\rangle\langle E_{k}|.
\end{equation}
 Here $\big\{|E_{k}\rangle\big\}$ are the eigenstates of the full
isolated many body system, and the probabilities $P_{k}$ are determined
by the initial state. Generally, $\bar{\boldsymbol{\rho}}_{\text{\textsc{d}}}$
is not identical to the microcanonical ensemble $\hat{\boldsymbol{\rho}}_{\text{mc}}$,
but many numerical studies show that, for a wide class of few body
observables $\hat{o}$, these two states give almost identical expectations,
i.e., $\mathrm{tr}[\bar{\boldsymbol{\rho}}_{\text{\textsc{d}}}\hat{o}]\simeq\mathrm{tr}[\hat{\boldsymbol{\rho}}_{\text{mc}}\hat{o}]$
(for nonintegrable systems). In this sense, effectively this isolated
system is said to achieve thermalization \citep{rigol_breakdown_2009,buca_unified_2023}.

For integrable systems, $\bar{\boldsymbol{\rho}}_{\text{\textsc{d}}}$
does not coincide with the microcanonical ensemble $\hat{\boldsymbol{\rho}}_{\text{mc}}$,
but with a generalized Gibbs ensemble $\hat{\boldsymbol{\rho}}_{\text{\textsc{gge}}}\sim\exp(-\sum\,\lambda_{m}\,\hat{\mathcal{I}}_{m})$
\citep{rigol_relaxation_2007}, where $\{\hat{\mathcal{I}}_{m}\}$
is the full set of the integrals of motion. In disordered systems
with many body localization \citep{oganesyan_localization_2007,pal_many-body_2010,altman_many-body_2018,abanin_colloquium_2019,znidaric_many-body_2008},
as well as some other systems with nonergodic properties \citep{wouters_quenching_2014,pozsgay_correlations_2014,turner_weak_2018,schecter_weak_2019,sala_ergodicity_2020,serbyn_quantum_2021,desaules_prominent_2023,lydzba_generalized_2023},
features seemingly defying ETH have been found \citep{rigol_breakdown_2009,buca_unified_2023}.

It is worth noting that the long-time averaging plays an essential
role in the above thermalization definition. For the dynamical relaxation
process, it remains desirable to find out a quantity that is capable
to to describe the irreversible entropy increase as in the standard
thermodynamics, and to understand the transition how the deterministic
reversible evolutions in few body systems become irreversible relaxations
with the increase of the system size. 

For this purpose, in analogy to Boltzmann's discussions on the isolate
gas \citep{boltzmann_weitere_1872,huang_statistical_1987}, we focus
on the time dependent evolution of the one-body states and the associated
observable expectations (\emph{local relaxation} \citep{hanggi_reaction-rate_1990,zwanzig_nonequilibrium_2001,cramer_exploring_2008,cramer_exact_2008,flesch_probing_2008}),
and study their relaxation processes in a finite chain of $N$ three-level
systems. We find that, though the total $N$-body state experiences
the deterministic unitary evolution, the local observable expectations
exhibit irreversible relaxation behaviors, namely, they first approach
certain values as their steady states and then experience small fluctuations
around them \citep{calabrese_time_2006}. Besides, for a finite size,
hierarchy recurrences (revivals) appear periodically, which can be
explained by the propagation of local excitation patterns \citep{cardy_thermalization_2014,li_hierarchy_2021,kang_correlational_2023,igloi_long-range_2000}.

Though the entropy of the full $N$-body state keeps a constant, we
find that the \emph{total correlation entropy} \citep{watanabe_information_1960,groisman_quantum_2005,zhou_irreducible_2008,anza_logarithmic_2020}
in this system approximately exhibits a monotonically increasing behavior
similar to that in the standard thermodynamics \citep{esposito_entropy_2010,ptaszynski_entropy_2019,ptaszynski_quantum_2023,manzano_entropy_2016,li_production_2017,you_entropy_2018,li_correlation_2019,li_hierarchy_2021,kang_correlational_2023}.
We calculate the possible maximum of the total correlation entropy
by variation under proper constraints, and find that the theoretical
maximum is achieved when each individual site is in a generalized
one-body Boltzmann state (GOBBS), i.e., $\tilde{\varrho}_{n}\sim\exp(-\beta_{\text{\textsc{e}}}\hat{H}_{n}-\beta_{\text{\textsc{s}}}\hat{S}_{n}^{z})$
(here $\beta_{\text{\textsc{e,s}}}$ are determined from the initial
state, $\hat{H}_{n}$, $\hat{S}_{n}^{z}$ are one body energy and
spin operators). And it turns out this theoretical correlation maximum
is highly coincident with the final increasing destination obtained
from the numerical result of the time dependent evolution. In this
sense, the total correlation entropy well indicates the dynamical
irreversibility of the relaxation process, which serves as an analogue
to the entropy increase in the standard thermodynamics. 

\begin{figure}
\includegraphics[width=1\columnwidth]{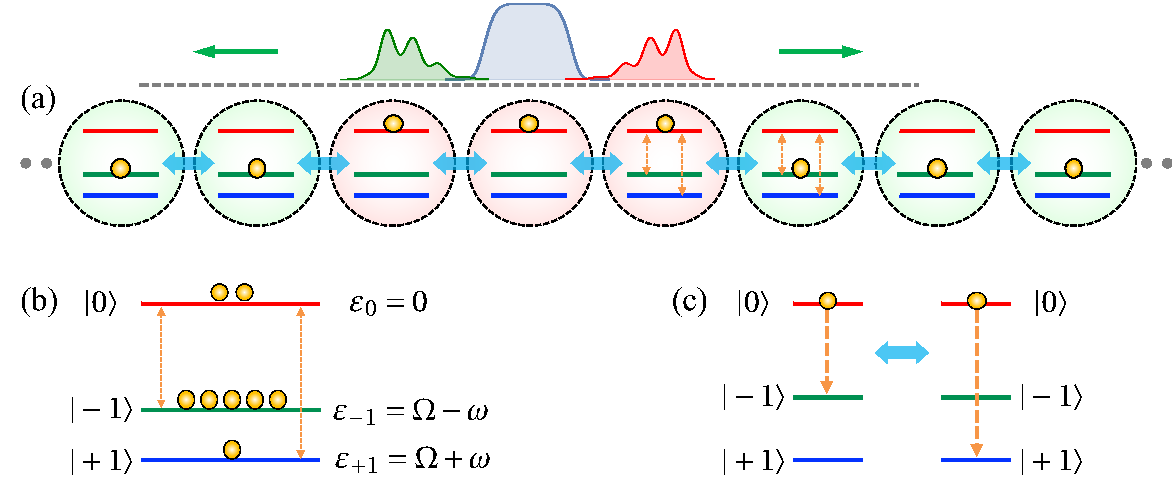}

\caption{(a) Schematics of a periodic chain consisting of $N$ interacting
three-level systems. (b) Energy levels for the onsite Hamiltonian
$\hat{H}_{n}$, which depend on $\omega$ and $\Omega$ (in this demonstration
$\Omega<\omega<0$, $\varepsilon_{+}<\varepsilon_{-1}<\varepsilon_{0}$).
(c) A demonstration for the interaction transition between two sites,
where the total magnetization is conserved, but not the onsite energy. }

\label{fig-chain}
\end{figure}

\vspace{0.5em}\noindent \emph{Interacting three-level systems} -
To illustrate our general idea, we consider the diffusion process
in a many body system consisting of an array of \noun{$N$ }three-level
systems with the periodic boundary condition, and they exchange energy
with the nearest neighbors (Fig.\,\ref{fig-chain}). Each three-level
system can be regarded as a \emph{pseudo-spin} (spin-1), and the full
$N$-body system is described by $\hat{\mathcal{H}}=\sum_{n=1}^{N}\hat{H}_{n}+\hat{H}_{\mathrm{int}}$,
where 
\begin{align}
\hat{H}_{n} & =\Omega\,(\hat{S}_{n}^{z})^{2}+\omega\,\hat{S}_{n}^{z}=\sum_{a=0,\pm1}\,\varepsilon_{n,a}\,|a\rangle_{n}\langle a|,\nonumber \\
\hat{H}_{\mathrm{int}} & =\sum_{n}J\,(\hat{S}_{n}^{+}\hat{S}_{n+1}^{-}+\hat{S}_{n}^{-}\hat{S}_{n+1}^{+})\label{eq:H}
\end{align}
are the onsite and interaction Hamiltonians respectively. Here $\hat{S}_{n}^{z}$,
$\hat{S}_{n}^{\pm}\equiv\hat{S}_{n}^{x}\pm i\hat{S}_{n}^{y}$  are
spin-1 operators on site-$n$, with $|\pm1\rangle_{n}$ and $|0\rangle_{n}$
the eigenstates of $\hat{S}_{n}^{z}$. The onsite energy levels of
site-$n$ are $\varepsilon_{n,0}=0$ and $\varepsilon_{n,\pm1}=\Omega\pm\omega$.
Except for few specific parameter sets, generally this model is nonintegrable
\citep{mutter_solvable_1995}.

It is easy to see that the total magnetization $\hat{\mathbf{S}}_{z}\equiv\sum_{n}\,\hat{S}_{n}^{z}$
is conserved ($[\hat{\mathbf{S}}_{z},\,\hat{\mathcal{H}}]=0$), thus
the full Hilbert space can be divided into subspaces labeled by different
\emph{magnon} numbers. For instance, setting the fully polarized state
$|\Theta\rangle:=\bigotimes_{n}|-1\rangle_{n}$ as the reference one,
the 1-magnon state $|\phi_{\alpha}\rangle:=\frac{1}{\sqrt{2}}\hat{S}_{\alpha}^{+}\,|\Theta\rangle$
indicates a local excitation generated at site-$\alpha$, and all
the $N$ states $\{|\phi_{\alpha}\rangle\}$ have the same total magnetization
$\mathsf{S}_{z}=1-N$, which span the \emph{1-magnon }subspace. Similarly,
$|\phi_{\alpha,\beta}\rangle\sim\hat{S}_{\alpha}^{+}\hat{S}_{\beta}^{+}\,|\Theta\rangle$
and $|\phi_{\alpha,\beta,\gamma}\rangle\sim\hat{S}_{\alpha}^{+}\hat{S}_{\beta}^{+}\hat{S}_{\gamma}^{+}\,|\Theta\rangle$
represent local magnon states generated on two or three local sites\footnote{The
indices $\alpha,\beta,\gamma$ are not necessarily different from
each other. But for 3-magnon states, $\alpha=\beta=\gamma$ should
be excluded since $(\hat{S}_{\alpha}^{+})^{3}=0$. Different permutations
of $\alpha,\beta,\gamma$ are equivalent, e.g., $|\phi_{\alpha,\beta}\rangle$
and $|\phi_{\beta,\alpha}\rangle$ give the same state. }, and they
span the\emph{ 2-magnon }and \emph{3-magnon} subspaces with $\mathsf{S}_{z}=2-N$
and $\mathsf{S}_{z}=3-N$ respectively. The subspace of \emph{n }magnons
also can be built in the same way \citep{wu_exact_2022,li_ground-state_2022,li_few-magnon_2024}. 

Though the full Hamiltonian (\ref{eq:H}) of $3^{N}$ dimension is
complicated to be solved, efficient numerical simulations of the magnon
dynamics can be achieved inside the \emph{n}-magnon subspace for a
small \emph{n} (see Appendix A). For example, the dimension for the
3-magnon subspace is $\frac{1}{6}N(N-1)(N+4)$ \citep{wu_exact_2022,li_ground-state_2022,li_few-magnon_2024}.
If the initial state of the system is chosen in the $n$-magnon subspace,
the evolution process is well constrained within this subspace, and
the full system state at any time can be obtained exactly.

\begin{figure}
\includegraphics[width=1\columnwidth]{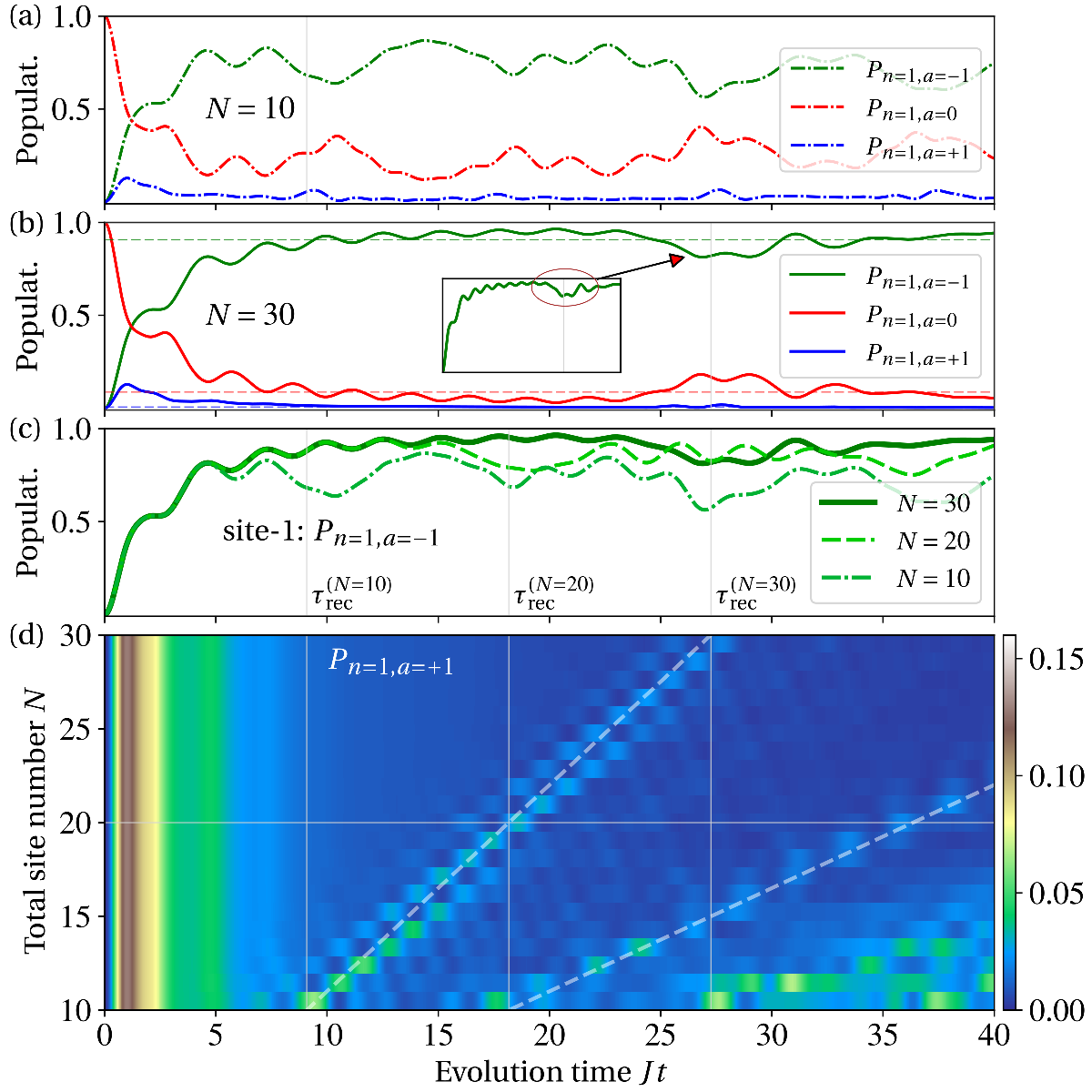}

\caption{(a, b) The time evolution of the populations on site-1 $P_{n=1,\,a=0,\pm1}(t)$
for chain sizes $N=10,\,30$. The dashed horizontal lines in (b) are
obtained by the long time average of $P_{n=1,a}(t)$ so as to estimate
their steady values in the fluctuating region. (c) The comparison
of $P_{n=1,a=-1}$ for $N=10,\,20,\,30$. (d) The comparison of $P_{n=1,a=+1}$
for $N=10,11,\dots,30$. The model parameters are set as $J=0.3$,
$\Omega=-1$, $\omega=-0.13$, which makes $|+1\rangle_{n}$ as the
onsite ground state with $\varepsilon_{+1}<\varepsilon_{-1}<\varepsilon_{0}$
(here $|\Omega|\equiv1$ is set as the energy unit, and the evolution
time is measured by $Jt$). The recurrence times are $\tau_{\mathrm{rec}}^{(N=10)}=\tau_{\mathrm{rec}}^{(N=20)}/2=\tau_{\mathrm{rec}}^{(N=30)}/3\simeq9.086\,J^{-1}$,
which are estimated from the propagation group velocity $\tau_{\mathrm{rec}}^{(N)}:=N/v_{\mathrm{g}}$. }

\label{fig-scale}
\end{figure}

\vspace{0.5em}\noindent \emph{Diffusion of local patterns} - We assume
initially the system is prepared in a localized 3-magnon state $|\Psi_{0}\rangle\sim\hat{S}_{1}^{+}\hat{S}_{2}^{+}\hat{S}_{3}^{+}|\Theta\rangle$
{[}Fig.\,\ref{fig-chain}(a){]}, and study the time dependent diffusion
process in this $N$-body system. The dynamics of the populations
$P_{n=1,a}(t)$ on site-1 is shown in Fig.\,\ref{fig-scale} and
compared for different chain sizes $N$. At first sight, the relaxation
process hardly exhibits any apparent regularity, except some seemingly
``random'' fluctuations. But when these evolution behaviors for
difference chain sizes $N$ are compared together {[}Fig.\,\ref{fig-scale}(c,
d){]}, we observe that their early-stage dynamics are almost identical
to each other before a certain \emph{recurrence time} $\tau_{\mathrm{rec}}^{(N)}$
{[}see the vertically parallel pattern to the left of the dashed white
slope in Fig.\,\ref{fig-scale}(d){]}.

Based on the above features of the onsite population dynamics {[}Fig.\,\ref{fig-scale}(a-c){]},
we can generally divide the relaxation process into three stages: 

{
\begin{enumerate}
\item \emph{Relaxation region} ($t<\tau_{\mathrm{rec}}^{(N)}$): the onsite
populations seem relaxing towards certain steady values \citep{calabrese_time_2006},
accompanied with some fluctuations;
\item \emph{Recurrence region} ($t\sim\tau_{\mathrm{rec}}^{(N)}$): a recurrence
``bump'' appears around a recurrence time $\tau_{\mathrm{rec}}^{(N)}$
{[}see the inset in Fig.\,\ref{fig-scale}(b){]}, which is proportional
to the chain size $N$;
\item \emph{Fluctuation region} ($t>\tau_{\mathrm{rec}}^{(N)}$): the onsite
populations fluctuate around certain central values, similar to a
homogenous stochastic process. In addition, hierarchy recurrences
appear around $t\sim\mathtt{q}\cdot\tau_{\mathrm{rec}}^{(N)}$ with
$\mathtt{q}=1,2,\dots$ {[}the dashed white slopes in Fig.\,\ref{fig-scale}(d){]}.
\end{enumerate}
}

\begin{figure}
\includegraphics[width=1\columnwidth]{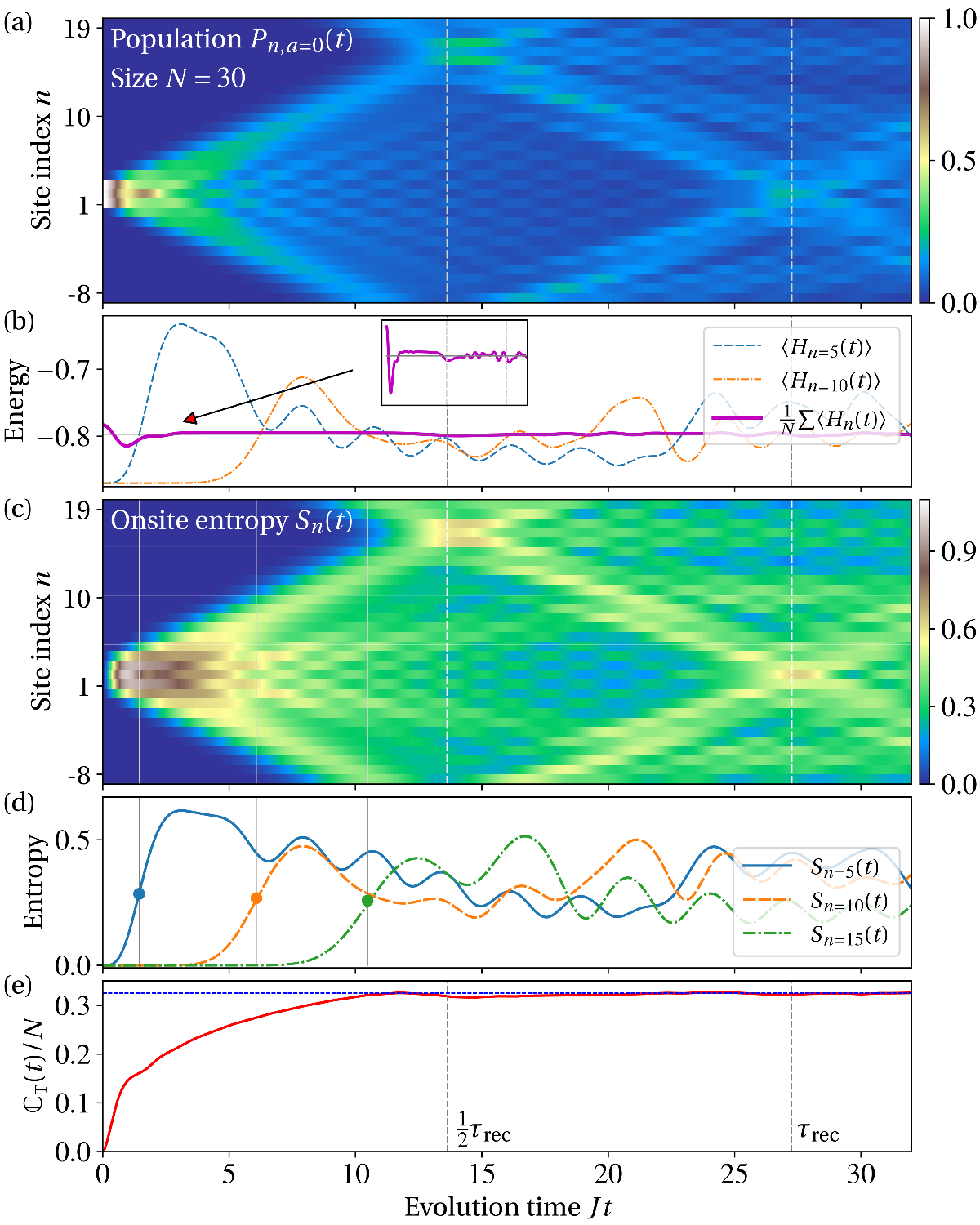}

\caption{(a) The diffusion pattern of onsite populations $P_{n,a=0}(t)$. (b)
The evolution of the average onsite energy $\sum\langle\hat{H}_{n}(t)\rangle/N$
comparing with the onsite energies $\langle\hat{H}_{n=5,10}(t)\rangle$.
The horizontal gray line is calculated from the long time average
of $\sum\langle\hat{H}_{n}(t)\rangle/N$. (c, d) The entropy dynamics
$S_{n}(t)$ on each site. $S_{n=5,10,15}(t)$ in (d) corresponds to
the three horizontal white lines in (c). The three dots in (d) are
the fastest increasing point (numerically fitted). (e) The evolution
of the total correlation entropy $\mathbb{C}_{\text{\textsc{t}}}(t)$,
comparing with the theoretical maximum (the horizontal dashed blue
line). The chain size is chosen as $N=30$, and all the other parameters
are the same as in Fig.\,\ref{fig-scale}. }

\label{fig-prop}
\end{figure}

Such relaxation and recurrence behaviors can be explained by the propagation
of the local excitations as follows {[}see Fig.\,\ref{fig-chain}(a)
and Fig.\,\ref{fig-prop}(a, c){]} \citep{cardy_thermalization_2014,li_hierarchy_2021,kang_correlational_2023}: 

Starting from the initial local excitations at site-1,2,3, the onsite
population patterns $P_{n,a}(t)$ propagate and diffuse to the two
sides of the chain. In the regime $t<\tau_{\mathrm{rec}}^{(N)}/2$,
the patterns diffuse with a constant speed, i.e., the Lieb-Robinson
group velocity \citep{lieb_finite_1972}. Because of the finite chain
size $N$, the two propagating patterns would meet each other at the
other side of the periodic chain around the site $n\sim N/2$, and
then propagate back to the initial starting sites, where they are
superposed together. This is just the moment when a recurrence bump
appears in the relaxation process of $P_{n=1,a}(t)$ {[}in Fig.\,\ref{fig-scale}(b){]}.
Such diffusion of the propagating patterns continues and gets superposed
again and again, which results in the hierarchy recurrences appearing
periodically around $t\sim\mathtt{q}\cdot\tau_{\mathrm{rec}}^{(N)}$.

The above observations suggest us defining the recurrence time as
$\tau_{\mathrm{rec}}^{(N)}:=N/v_{\mathrm{g}}$, where $v_{\mathrm{g}}$
is the group velocity of the propagating patterns in the diffusion
process. Here $v_{\mathrm{g}}\equiv\Delta n/\Delta t$ is numerically
fitted by checking the moments that the propagating pattern of the
onsite entropy $S_{n}\equiv-\mathrm{tr}[\hat{\rho}_{n}\ln\hat{\rho}_{n}]$
reaches site-$n$ {[}$\hat{\rho}_{n}(t)$ is the reduced density state
of site-$n${]}, and the reaching time on site-$n$ is defined as
the fastest increasing moment of $S_{n}(t)$ {[}e.g., see the three
dots for site-5,10,15 in Fig.\,\ref{fig-prop}(d), corresponding
to the lines in Fig.\,\ref{fig-prop}(c){]}.

The linear dependence of the recurrence time $\tau_{\mathrm{rec}}^{(N)}$
on the chain size $N$ {[}Fig.\,\ref{fig-scale}(d){]} confirms that
the diffusion speed $v_{\mathrm{g}}$ remains the same for different
system sizes, which is consistent with the diffusion in the macroscopic
world. For instance, comparing the diffusion of an ink drop in a glass
of water with that in a huge pool, the diffusion speeds should be
the same, but definitely it costs much longer time for the ink drop
to fully diffuse all over the bigger volume. 

\vspace{0.5em}\noindent  \emph{Dynamical irreversibility }- The above
dynamical behaviors of the pattern propagations well exhibit how the
macroscopic diffusion behavior emerge with the increase of the system
size $N$. In the thermodynamic limit $N\rightarrow\infty$, within
a finite observation time $(t\ll\tau_{\mathrm{rec}}\sim\infty$),
we would be convinced that the system is relaxing towards a certain
steady state irreversibly. But here we still need a physical quantity
to describe such dynamical irreversibility, which corresponds to the
entropy increase in the standard macroscopic thermodynamics.

For such a time-dependent nonequilibrium process, the temperature
is not well defined, thus the concept of thermal entropy $\text{đ}Q/T$
does not apply either. Since the full system follows the unitary evolution,
the von Neumann entropy of the full $N$-body state $\hat{\boldsymbol{\rho}}(t)$
always keeps a constant as the initial one \citep{huang_statistical_1987,hobson_irreversibility_1966,mackey_dynamic_1989,landi_irreversible_2021}.
On the other hand, the onsite entropy of each individual site fluctuates
from time to time {[}Fig.\,\ref{fig-prop}(c, d){]}, thus cannot
be used as the indicator for the above irreversibility in this nonequilibrium
process either.

Instead, we suggest using the total correlation entropy to describe
the aforementioned dynamical irreversibility, which is defined as
\citep{watanabe_information_1960,groisman_quantum_2005,zhou_irreducible_2008,anza_logarithmic_2020}
\begin{equation}
\mathbb{C}_{\text{\textsc{t}}}[\hat{\boldsymbol{\rho}}]:=\sum_{n=1}^{N}S_{\text{\textsc{v}}}[\hat{\rho}_{n}]-S_{\text{\textsc{v}}}[\hat{\boldsymbol{\rho}}].
\end{equation}
 Here $S_{\text{\textsc{v}}}[\hat{\rho}]\equiv-\mathrm{tr}[\hat{\rho}\ln\hat{\rho}]$
is the von Neumann entropy, and $\hat{\rho}_{n}$ are the reduced
one-body states of the full $N$-body state $\hat{\boldsymbol{\rho}}$.
$\mathbb{C}_{\text{\textsc{t}}}[\hat{\boldsymbol{\rho}}]$ measures
the total amount of correlations inside the $N$-body system \citep{watanabe_information_1960,zhou_irreducible_2008}.
For two body systems ($N=2)$, $\mathbb{C}_{\text{\textsc{t}}}[\hat{\boldsymbol{\rho}}]$
is reduced to the mutual information, which is widely used as a measure
for the bipartite correlation.

It turns out that the total correlation entropy $\mathbb{C}_{\text{\textsc{t}}}[\hat{\boldsymbol{\rho}}(t)]$
roughly exhibits a monotonically increasing behavior during the diffusion
process {[}Fig.\,\ref{fig-prop}(e){]}. Around the time $t\sim\tau_{\mathrm{rec}}/2$,
the total correlation reaches its maximum, and hereafter almost keeps
this maximal value, accompanied by only quite small fluctuations.
Such behaviors are well consistent with the irreversible entropy increase
in the standard macroscopic thermodynamics.

Based on such an increasing behavior, now we ask: what is the possible
maximum of the total correlation after a long time growth? To find
out this theoretical maximum, we consider the variation of $\mathbb{C}_{\text{\textsc{t}}}[\{p_{n,a}\}]=-\sum_{n,a}\,p_{n,a}\ln p_{n,a}$
over all possible population configurations $\{p_{n,a}\}$ for each
onsite levels. Here some constraint conditions should be taken into
account: (1) probability normalization $p_{n,+1}+p_{n,0}+p_{n,+1}=1$,
(2) total magnetization conservation $\sum_{n}\,(p_{n,+1}-p_{n,-1})=\mathsf{S}_{z}$.
In addition, though not exactly conserved, it is reasonable to assume
the total onsite energy $\sum_{n}\,\langle\hat{H}_{n}(t)\rangle$
would approximately remain the same as the initial one, especially
when the interaction strength $J$ is small {[}see Fig.\,\ref{fig-prop}(b){]},
and that gives another constraint (3) $\sum_{n,a}\,\varepsilon_{n,a}\,p_{n,a}\simeq\mathcal{E}_{0}$. 

With the help of Lagrangian multipliers \citep{jaynes_information_1957,jaynes_gibbs_1965},
it can be proved that the theoretical maximum of the total correlation
is achieved when each individual site takes the state (see Appendix
B)
\begin{equation}
\tilde{\varrho}_{n}\sim\exp(-\beta_{\text{\textsc{e}}}\,\hat{H}_{n}-\beta_{\text{\textsc{s}}}\,\hat{S}_{n}^{z}),\label{eq:rho_eq}
\end{equation}
where $\beta_{\text{\textsc{e,s}}}$ are two Lagrangian multipliers
shared by all the $N$ sites, and they can be determined from the
above constraints and the initial state. Clearly this state has a
form of a generalized Boltzmann distribution \citep{goldstein_canonical_2006,popescu_entanglement_2006},
thus we call it a generalized one-body Boltzmann state (GOBBS).

For the example of the parameters in Fig.\,\ref{fig-scale}, the
total onsite energy from the initial state gives $\mathcal{E}_{0}=(N-3)\varepsilon_{-1}$
with $N=30$, and it turns out the above theoretical maximum from
variation is achieved when all the sites take the same distribution
$\tilde{p}_{n,-1}=0.9$, $\tilde{p}_{n,0}=0.1$, $\tilde{p}_{n,+1}=0$,
which gives the possible maximum $\tilde{\mathbb{C}}_{\text{\textsc{t}}}/N\simeq0.3251$
{[}dashed blue line in Fig.\,\ref{fig-prop}(e){]}. It is worth noting
that in this example the three energy levels satisfy $\varepsilon_{+1}<\varepsilon_{-1}<\varepsilon_{0}$,
showing that such a distribution clearly does not obey the standard
Boltzmann one that solely depends on the energies exponentially.

Now we compare this theoretical maximum with the numerical results
obtained from the exact time evolution. The total onsite energy does
remain similar to the initial value {[}purple line in Fig.\,\ref{fig-prop}(b){]},
and shows much weaker fluctuation than each individual sites, which
guarantees the constraint (3) is reliable enough. In the fluctuation
region ($t>\tau_{\mathrm{rec}}$), the local observable expectations
(e.g., the onsite populations, energy) fluctuate around certain central
values, which can be regarded as the ``steady'' states of the relaxation
process. Thus, long time averages could eliminate these fluctuations
and give a reasonable estimation for these steady values (see Appendix
D). In this sense, under the parameters in Fig.\,\ref{fig-scale},
the steady values of onsite populations are obtained as $p_{n,-1}^{\infty}\simeq0.9069$,
$p_{n,0}^{\infty}\simeq0.0862$, $p_{n,+1}^{\infty}\simeq0.0069$
{[}dashed horizontal lines in Fig.\,\ref{fig-scale}(b){]}, which
give the steady total correlation as $\mathbb{C}_{\text{\textsc{t}}}^{\infty}/N\simeq0.3342$,
and this turn out to be highly coincident with the theoretical maximum
$\tilde{\mathbb{C}}_{\text{\textsc{t}}}/N\simeq0.3251$ {[}dashed
blue line in Fig.\,\ref{fig-prop}(e){]}.

\vspace{0.5em}\noindent \emph{Summary} - In this paper, we study
the relaxation dynamics in an isolated many body system. Though the
whole system experience the reversible unitary evolution, the dynamics
of the local states exhibit irreversible relaxation behaviors; the
entropy of the full many body state keeps unchanged, while the total
correlation entropy exhibits a monotonically increasing behavior,
which well indicates the dynamical irreversibility of the relaxation
process \citep{esposito_entropy_2010,ptaszynski_entropy_2019,ptaszynski_quantum_2023,manzano_entropy_2016,li_production_2017,you_entropy_2018,li_correlation_2019,li_hierarchy_2021,kang_correlational_2023}.
Indeed, most macroscopic thermodynamic observables in practice only
involve the multi-particle average of local observables (e.g., the
gas temperature and pressure are determined by the average energy
of single molecular). Therefore, based on the similar idea, here we
only focus on local states and observable expectations, but do not
concern whether the full $N$-body state could reach the microcanonical
or Gibbs state.

More importantly, we find that the total correlation entropy approximately
exhibits a monotonically increasing behavior. Moreover, the variation
analysis shows that, the theoretical correlation maximum would be
achieved when each individual site stays in a GOBBS which is no longer
solely determined by energies, and surprisingly this theoretical maximum
is highly coincident with the the exact time dependent evolution.
Namely, the local sites appear approaching GOBBS as their steady states
irreversibly. In this sense, the entropy of the full state keeps unchanged,
while the total correlation entropy exhibit the irreversible growth,
which serves as a potential generalization for the irreversible entropy
production in the standard thermodynamics.

\vspace{0.5em}\noindent \emph{Acknowledgments }- SWL appreciates
quite much for the helpful discussion with Z. H. Wang and L.-P. Yang.
NW is supported by the National Key Research and Development Program
of China under Grant No. 2021YFA1400803.

\appendix

\section{Matrix elements of the Hamiltonian in the 3-magnon subspace}

Here we write down the matrix elements of the system Hamiltonian in
the local magnon basis. Since the magnon number is conserved in this
system, to obtain the eigenstates of the $N$-body Hamiltonian, the
diagonalization process can be constrained in the $n$-magnon subspace. 

Focusing on the 3-magnon subspace spanned by $|\phi_{\alpha,\beta,\gamma}\rangle\sim\hat{S}_{\alpha}^{+}\hat{S}_{\beta}^{+}\hat{S}_{\gamma}^{+}\,|\Theta\rangle$
with $|\Theta\rangle:=\bigotimes_{n}|-1\rangle_{n}$, the matrix elements
of the system Hamiltonian are given by $\langle\phi_{\alpha'\beta'\gamma'}|\hat{\mathcal{H}}|\phi_{\alpha\beta\gamma}\rangle$.
Since different permutation orders of $\alpha,\beta,\gamma$ are equivalent,
when arranged in the ascending order $\alpha\le\beta\le\gamma$, the
basis states $|\phi_{\alpha\beta\gamma}\rangle$ can be divided into
two types ($\alpha,\beta,\gamma$ cannot be equal to each other at
the same time): 
\begin{enumerate}
\item $1\le\alpha<\beta<\gamma\le N$, totally $C_{N}^{3}$ states; 
\item $1\le\alpha=\beta<\gamma\le N$ and $1\le\alpha<\beta=\gamma\le N$,
totally $2C_{N}^{2}$ states. 
\end{enumerate}
Thus the dimension of the 3-magnon subspace is $\frac{1}{6}N(N-1)(N+4)$. 

Now we write down the matrix elements of the operators under the basis
$|\phi_{\alpha\beta\gamma}\rangle$. For $\hat{S}_{n}^{z}$, notice
that $\langle\phi_{\alpha'\beta'\gamma'}|\hat{S}_{n}^{z}|\phi_{\alpha\beta\gamma}\rangle=\delta_{\alpha\alpha'}\delta_{\beta\beta'}\delta_{\gamma\gamma'}\,\langle\phi_{\alpha\beta\gamma}|\hat{S}_{n}^{z}|\phi_{\alpha\beta\gamma}\rangle$
is diagonal, and these matrix elements can be summarized as (here
the order in the indices is irrelevant) 
\begin{equation}
\begin{cases}
\langle\phi_{\alpha,\beta,\gamma}|\hat{S}_{n}^{z}|\phi_{\alpha,\beta,\gamma}\rangle=-1, & n\neq\alpha,\beta,\gamma\\
\langle\phi_{n,\beta,\gamma}|\hat{S}_{n}^{z}|\phi_{n,\beta,\gamma}\rangle=0, & n\neq\beta,\gamma\ \text{and }\beta\neq\gamma\\
\langle\phi_{n,n,\gamma}|\hat{S}_{n}^{z}|\phi_{n,n,\gamma}\rangle=1, & n\neq\gamma
\end{cases}
\end{equation}
In the first line, $\alpha,\beta,\gamma$ are not necessarily to be
equal or unequal to each other.

For the interaction terms $\langle\phi_{\alpha'\beta'\gamma'}|\hat{S}_{n}^{+}\hat{S}_{n+1}^{-}|\phi_{\alpha\beta\gamma}\rangle$,
the nonzero elements require that $\hat{S}_{n}^{+}\hat{S}_{n+1}^{-}|\phi_{\alpha\beta\gamma}\rangle$
must be still in the 3-magnon subspace, thus at least one of $\alpha,\beta,\gamma$
must be $n+1$, otherwise $\hat{S}_{n+1}^{-}|\phi_{\alpha\beta\gamma}\rangle=0$.
As a result, the nonzero elements of $\langle\phi_{\alpha'\beta'\gamma'}|\hat{S}_{n}^{+}\hat{S}_{n+1}^{-}|\phi_{\alpha\beta\gamma}\rangle$
can be summarized as (here the order in the indices is irrelevant)
\begin{equation}
\langle\phi_{n,\beta,\gamma}|\hat{S}_{n}^{+}\hat{S}_{n+1}^{-}|\phi_{n+1,\beta,\gamma}\rangle=2,
\end{equation}
where $\beta,\gamma\in\{1,\dots,N\}$ , but cannot take $n+1$ or
$n$ at the same time. 

Similarly, when considering the long range interaction, the nonzero
elements of $\hat{S}_{n}^{+}\hat{S}_{n+d}^{-}$ are $\langle\phi_{n,\beta,\gamma}|\hat{S}_{n}^{+}\hat{S}_{n+d}^{-}|\phi_{n+d,\beta,\gamma}\rangle=2$,
where $\beta,\gamma\in\{1,\dots,N\}$, but cannot take $n+d$ or $n$
at the same time. 

\section{Correlation maximization }

Here we show how the theoretical maximum of the total correlation
is obtained with the help of the Lagrangian multipliers. During the
evolution, the density state of each individual site keeps diagonal,
$\hat{\varrho}_{n}(t)=\sum_{a}\,p_{n,a}(t)\,|a\rangle_{n}\langle a|$,
and we need to consider three constraint conditions: (1) probability
normalization $p_{n,+1}+p_{n,0}+p_{n,+1}=1$, (2) total magnetization
conservation $\sum_{n}\,(p_{n,+1}-p_{n,-1})=\mathsf{S}_{z}$, (3)
the approximate conservation of total onsite energy $\sum_{n,a}\,\varepsilon_{n,a}p_{n,a}\simeq\mathcal{E}_{0}$.
Thus, to find out the possible maximum of the total correlation entropy
among all possible population configurations $\{p_{n,a}\}$, we study
the variation on the following quantity, 
\begin{align}
\mathcal{T}:= & -\sum_{n,a}p_{n,a}\ln p_{n,a}-\beta_{\text{\textsc{s}}}\Big[\mathsf{S}_{z}-\sum_{n}(p_{n,+1}-p_{n,-1})\Big]\nonumber \\
- & \beta_{\text{\textsc{e}}}\Big[\mathcal{E}_{0}-\sum_{n,a}\varepsilon_{n,a}\,p_{n,a}\Big]-\sum_{n}\lambda_{n}(1-\sum_{a}p_{n,a}).
\end{align}
Here $\beta_{\text{\textsc{e,s}}}$, $\lambda_{n}$ are Lagrangian
multipliers, and the variation gives 
\begin{align}
\frac{\partial\mathcal{T}}{\partial p_{n,+1}} & =-\ln p_{n,+1}-1+\beta_{\text{\textsc{s}}}+\beta_{\text{\textsc{e}}}\varepsilon_{n,+1}+\lambda_{n}\equiv0,\nonumber \\
\frac{\partial\mathcal{T}}{\partial p_{n,-1}} & =-\ln p_{n,-1}-1-\beta_{\text{\textsc{s}}}+\beta_{\text{\textsc{e}}}\varepsilon_{n,-1}+\lambda_{n}\equiv0,\nonumber \\
\frac{\partial\mathcal{T}}{\partial p_{n,0}} & =-\ln p_{n,0}-1+\lambda_{2}\varepsilon_{n,0}+\lambda_{n}\equiv0.
\end{align}
These three equations further gives 
\begin{align}
\ln(p_{n,+1}/p_{n,0}) & =\beta_{\text{\textsc{e}}}(\varepsilon_{n,+1}-\varepsilon_{n,0})+\beta_{\text{\textsc{s}}},\nonumber \\
\ln(p_{n,-1}/p_{n,0}) & =\beta_{\text{\textsc{e}}}(\varepsilon_{n,-1}-\varepsilon_{n,0})-\beta_{\text{\textsc{s}}}.
\end{align}
The solution of these two equations can be written as 
\begin{equation}
\tilde{\varrho}_{n}\sim\exp(-\beta_{\text{\textsc{e}}}\,\hat{H}_{n}-\beta_{\text{\textsc{s}}}\,\hat{S}_{n}^{z}).
\end{equation}
Namely, the possible maximum of the total correlation entropy is achieved
if the reduced density matrix of each site take $\tilde{\varrho}_{n}$.
Clearly, such a state has a form of a generalized one-body Boltzmann
distribution, and all the $N$ sites share the same two parameters
$\beta_{\text{\textsc{e,s}}}$, which can be further determined from
the above constraints. 

For the example of the homogeneous case, since all $\hat{H}_{n}$
have the same form, all the $N$ sites have the same distribution,
i.e., $p_{n,a}\equiv\tilde{p}_{a}$, and the above constraints give
\begin{gather}
\tilde{p}_{+1}-\tilde{p}_{-1}=\mathsf{S}_{z}/N,\nonumber \\
\varepsilon_{+1}\,\tilde{p}_{+1}+\varepsilon_{0}\,\tilde{p}_{0}+\varepsilon_{-1}\,\tilde{p}_{-1}=\mathcal{E}_{0}/N,\\
\tilde{p}_{+1}+\tilde{p}_{0}+\tilde{p}_{-1}=1,\nonumber 
\end{gather}
 which gives the result shown in the main text.

\section{Connection with the Boltzmann gas in the standard thermodynamics }

For an isolated classical gas with weak collisions, the full ensemble
state $\boldsymbol{\rho}(\vec{P},\vec{Q},t)$ of the $N$-body system
follows the Liouville equation $\partial_{t}\boldsymbol{\rho}=-\{\,\boldsymbol{\rho},\,\mathcal{H}(\vec{P},\vec{Q})\,\}$
{[}here $(\vec{P},\vec{Q}):=(\vec{p}_{1},\vec{p}_{2},\dots;\vec{q}_{1},\vec{q}_{2},\dots)${]}.
As a result, the Gibbs entropy of the full $N$-body state, 
\begin{equation}
S_{\text{\textsc{g}}}[\boldsymbol{\rho}(\vec{P},\vec{Q},t)]=-\int d^{3N}p\,d^{3N}q\,\boldsymbol{\rho}\ln\boldsymbol{\rho},
\end{equation}
 never changes with time.

On the other hand, if we focus on the one-particle probability distribution
function (PDF) $\varrho(\mathbf{p}_{1},\mathbf{r}_{1},t)$, it can
be described by Boltzmann's transport equation (denoting $\mathrm{d}\boldsymbol{\varsigma}_{i}:=d^{3}\mathbf{r}_{i}\,d^{3}\mathbf{p}_{i}$)
\citep{li_correlation_2019}, 
\begin{align}
\big[\partial_{t}+ & \frac{\mathbf{p}_{1}}{m}\cdot\nabla_{\mathbf{r}_{1}}\big]\,\varrho(\mathbf{p}_{1},\mathbf{r}_{1},t)\nonumber \\
= & \int(\rho_{1'2'}-\rho_{12})\chi_{[12\rightarrow1'2']}\,\mathrm{d}\boldsymbol{\varsigma}_{1}'\,\mathrm{d}\boldsymbol{\varsigma}_{2}'\,\mathrm{d}\boldsymbol{\varsigma}_{2}.\label{eq:Boltz}
\end{align}
The LHS roots from the free motion of the single particle, while the
RHS comes from the two-body collision with the other particles. Here
$\chi_{[12\rightarrow1'2']}$ denotes the transition ratio from the
initial state $(\mathbf{p}_{1}\mathbf{r}_{1};\,\mathbf{p}_{2}\mathbf{r}_{2})$
scattered into the final state $(\mathbf{p}_{1}'\mathbf{r}_{1}';\,\mathbf{p}_{2}'\mathbf{r}_{2}')$,
and $\rho_{12}\equiv\rho(\mathbf{p}_{1}\mathbf{r}_{1};\,\mathbf{p}_{2}\mathbf{r}_{2},t)$,
$\rho_{1'2'}\equiv\rho(\mathbf{p}_{1}'\mathbf{r}_{1}';\,\mathbf{p}_{2}'\mathbf{r}_{2}',t)$
indicate the two-particle joint PDF. 

For further discussions, the \emph{molecular-disorder assumption}
is needed, i.e., the two-particle joint PDF approximately equals the
product of the two one-particle PDF $\rho(\mathbf{p}_{1}\mathbf{r}_{1};\mathbf{p}_{2}\mathbf{r}_{2},t)\simeq\varrho(\mathbf{p}_{1},\mathbf{r}_{1},t)\cdot\varrho(\mathbf{p}_{2},\mathbf{r}_{2},t)$.
That would further lead to the Boltzmann \emph{H}-theorem, i.e., it
turns out the entropy of the one-particle PDF,
\begin{equation}
S[\varrho(\mathbf{p},\mathbf{r},t)]=-\int d^{3}p\,d^{3}r\,\varrho\ln\varrho,
\end{equation}
 keeps increasing monotonically, until the one-particle PDF $\varrho(\mathbf{p},\mathbf{r},t)$
reaches the Boltzmann-Maxwell distribution $\varrho\sim\exp\big[-\beta_{\text{\textsc{e}}}\,H(\mathbf{p},\mathbf{r})\,\big]$,
where $H(\mathbf{p},\mathbf{r})=\mathbf{p}^{2}/2m+V(\mathbf{r})$
is the single particle Hamiltonian (the instant two-body collisions
are not omitted here). 

In sum, the entropy of the full $N$-body state follows the Liouville
dynamics and never changes, while the entropy from the one-particle
PDF exhibits the irreversible growth. In principle, the full $N$-body
distribution $\boldsymbol{\rho}(\vec{P},\vec{Q},t)$ would gradually
become a highly correlated state. However, generally the exact form
of $\boldsymbol{\rho}(\vec{P},\vec{Q},t)$ is not concerned, since
in practice the one-particle PDF $\varrho(\mathbf{p},\mathbf{r},t)$
is enough to give the expectations for most macroscopic thermodynamic
observables, such as the temperature, gas pressure, and some other
more general few-body correlation functions. 

Our study follows the similar idea as above. Clearly, here the total
correlation entropy $\mathbb{C}_{\text{\textsc{t}}}(t)=\sum_{n}\,S[\varrho(\mathbf{p}_{n},\mathbf{r}_{n},t)]-S_{\text{\textsc{g}}}[\boldsymbol{\rho}(\vec{P},\vec{Q},t)]$
well returns the irreversible entropy increase behavior in the standard
thermodynamics quantitatively. 

\section{The long time behavior of local observables}

In our study, we mainly focus on the time dependent evolution of the
expectations for local observables, such as the populations and onsite
energy of each individual site, which are calculated by 
\begin{align*}
\langle\hat{o}(t)\rangle & =\langle\Psi_{0}|\,e^{i\hat{H}t}\,\hat{o}\,e^{-i\hat{H}t}|\Psi_{0}\rangle\\
 & =\sum_{k,q}\langle E_{k}|\hat{o}|E_{q}\rangle\cdot\langle\Psi_{0}|E_{k}\rangle\langle E_{k}|\Psi_{0}\rangle\cdot e^{i(E_{k}-E_{q})t}.
\end{align*}
 From the numerical results shown in the main text, it can be noticed
that the dynamics of these local observable expectations generally
contains three stages: (1) relaxation region ($t\apprle\tau_{\mathrm{rec}}$),
(2) recurrence region ($t\sim\tau_{\mathrm{rec}}$), (3) fluctuation
region ($t\apprge\tau_{\mathrm{rec}}$). 

\begin{figure}
\includegraphics[width=0.95\columnwidth]{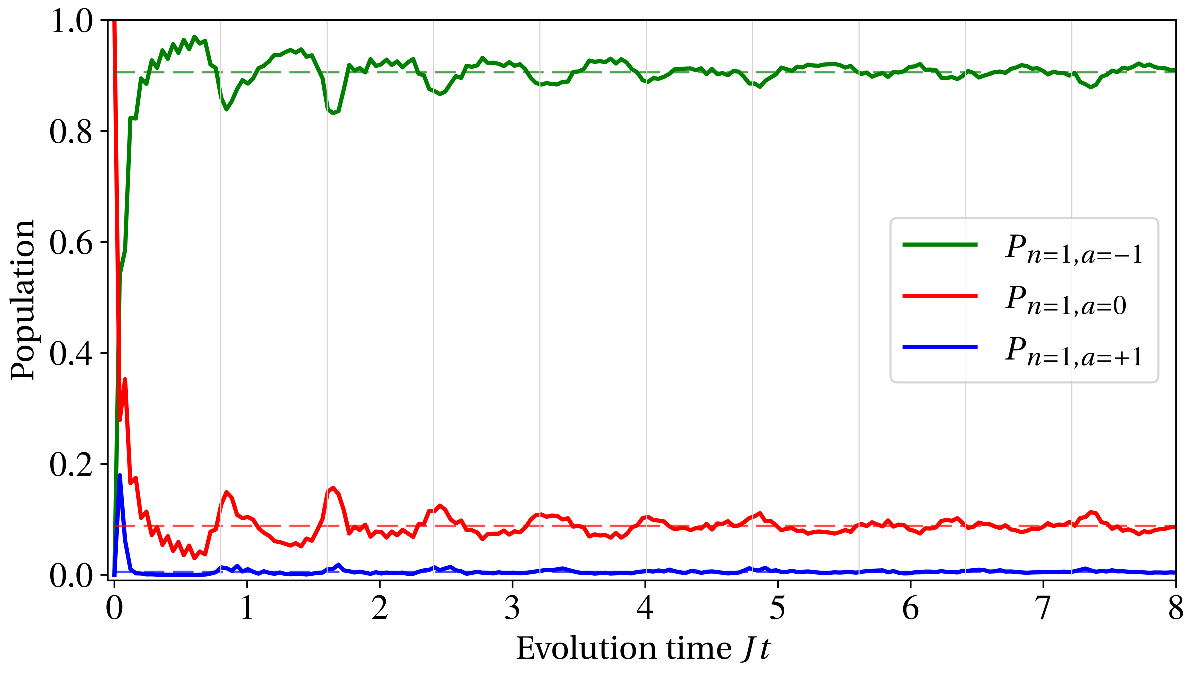}

\caption{The time dependent evolution of the populations on site-1 $P_{n=1,\,a=0,\pm1}$.
Here the model parameters are set as $N=30$, $J=10$, $\Omega=-1$,
$\omega=-0.13$. The recurrence time is estimated from the propagation
pattern of the onsite entropy as mentioned in the main text, which
gives $\tau_{\mathrm{rec}}\simeq0.8015\,J^{-1}$ (the solid vertical
gray lines). The dashed horizontal lines are calculated by the time
averages, i.e., $p_{n,a}^{\infty}$.}

\label{fig-long}
\end{figure}

In the fluctuation region, these local observable expectations $\langle\hat{o}(t)\rangle$
exhibit\textsf{ }some seemingly “random” and small fluctuations around
certain central values. In this sense, the central values of these
fluctuations can be regarded as the final destination of the relaxation
process. To find out the central values of these fluctuations, the
long time average makes a reasonable estimation by cancelling the
fluctuations, i.e.,

\[
o_{\infty}:=\lim_{T\rightarrow\infty}\frac{1}{T}\int_{0}^{T}dt\,\langle\hat{o}(t)\rangle=\sum_{k}\langle E_{k}|\hat{o}|E_{k}\rangle\left|\langle\Psi_{0}|E_{k}\rangle\right|^{2}.
\]

Such a treatment of long time average also has been widely adopted
in thermalization problems which were based on the idea of ergodicity.
In our study, the ``steady states'' of the onsite populations $p_{n,a}^{\infty}$
and the long time average of the total onsite energy are calculated
in this way. 

Fig.\,\ref{fig-long} shows the evolution for a long time. In the
fluctuation region long after the relaxation, the populations $P_{n,a}(t)$
exhibit\textsf{ }small fluctuations around certain central values,
and these fluctuation centers can be well estimated by the above time
average ($p_{n=1,a=-1}^{\infty}\simeq0.9057$, $p_{n=1,a=0}^{\infty}\simeq0.0886$,
$p_{n=1,a=-1}^{\infty}\simeq0.0057$). Besides, clearly the appearance
moments of the hierarchy recurrences well fit the relation $t\sim\mathtt{q}\cdot\tau_{\text{rec}}$
with $\mathtt{q}=1,2,\dots$ (see the solid vertical gray lines).

\end{document}